\documentclass[epj,referee]{svjour}
\usepackage{epsf}
\begin{document}
\title{
Numerical diagonalization analysis of the
ground-state superfluid-localization transition
in two dimensions
}
\titlerunning{Two-dimensional ground-state superfluid-localization transition}
\author{Yoshihiro Nishiyama}
\institute{
Department of Physics, Faculty of Science,
Okayama University,
Okayama 700-8530, Japan}
\date{Received: date / Revised version: date}
%
\abstract{Ground state of the
two-dimensional hard-core-boson system in the presence of the 
quenched random
chemical
potential is investigated by means of 
the exact-diagonalization method
for the system sizes up to $L=5$.
The criticality and the DC conductivity at the
superfluid-localization transition have been controversial so far.
We estimate, with the finite-size scaling analysis,
the correlation-length and the dynamical critical exponents as
$\nu=2.3\pm0.6$ and $z=2$, respectively.
The AC conductivity is computed with
the Gagliano-Balseiro formula,
with which the resolvent (dynamical response function)
is 
expressed in terms of the continued-fraction form
consisted of Lanczos tri-diagonal elements.
Thereby, we estimate the universal DC conductivity
as $\sigma_{\rm c}(\omega\to0)=0.135\pm0.01((2e)^2/h)$.
} 
\maketitle
\section{Introduction}
\label{section1}
The scaling argument of Abrahams, Anderson, Licciardello and Ramakrishnan
\cite{Abrahams79}
states that in {\em two} dimensions, infinitesimal amount of
quenched randomness should drive itinerant extended states
to localize.
That is, at the absolute zero temperature, the conductivity
should be vanishing, if there exist any randomnesses.
These are, however, some exceptions where the above description fails.
For instance, the integer quantum hall effect is
described in terms of {\em successive} metal-insulator 
(delocalization-localization) transitions
of dirty two-dimensional electron system
with the external magnetic field varied.
In the above-mentioned scaling theory,
the random perturbation is appeared to be marginal so that 
some unexpected factors, namely, the magnetic field and
the many-body interaction, would possibly change the 
scenario.

For instance, suppose that there exists an attractive interaction
among the electrons.
At the ground state, the electrons would be unstable against
the bose condensation so that the system would be in the
superconducting phase.
The disorder-driven localization from the superconducting
phase is apparently out of the scope of the conventional
localization theory, and has been studied extensively so far.
In experiments, the transition is observed for
metallic films \cite{Haviland89,Liu91,Lee90,Hebard85},
high-$T_{\rm c}$ films \cite{Wang91}
and Josephson-junction arrays \cite{Geerligs89,Katsumoto95}.
One of the main concerns is the
conductivity at the critical point:
The experiments show that irrespective of the samples examined,
the localization transition occurs at a universal condition.
Namely, at the transition, the electrical conductivity is
found to be $\approx(2e)^2/h$.
(The parameter $e$ denotes the charge of a single electron,
and $h$ denotes the Planck constant.)

Localization transition of the absorbed helium by a porous media 
\cite{Crooker83,Finotello88}
is considered \cite{Ma86,Fisher89} to belong 
precisely to the
same universality class mentioned above.
(That is, the critical exponents are identical.
The electrical conductivity, however, does not make any sense
for the latter,
because the helium atom is not charged.)
The equivalence is based on the belief that
the Cooper pair is formed already in the localization phase as well as
in the superconducting phase, and the
essence of the transition is concerned only in the boson degrees of
freedom and the site-random chemical potential.
This picture was found to be valid in one dimension 
\cite{Giamarch87}.

From a theoretical viewpoint,
the criticality itself is a matter of interest. 
It is notable
that the present phenomenon occurs at the ground state. 
This criticality is different essentially from that of
the finite-temperature
transitions.
In the path-integral picture for the partition function,
$d$-dimensional quantum system is regarded as a $(d+1)$-dimensional
classical system.
The system-size along the imaginal-time direction is given
by the inverse temperature, which is diverging
at the ground state.
Hence, the critical fluctuation along the imaginary-time direction
contributes to the universality class significantly.
This extra contribution is characterized by the
dynamical critical exponent $z$,
which is explained in the next section in detail.

The scaling argument \cite{Fisher89,Fisher90} shows that the conductivity
remains finite at the onset of the localization transition
even at the ground state.
Moreover, the argument for general
dimensions yields the dynamical critical exponent
being equal to the spacial dimension; $z=d$.
We explain the argument in the subsection \ref{section2_2}.
This prediction is astonishing in the sense that there is no
upper critical dimension in this critical phenomenon.
(The prediction is confirmed rigorously for $d=1$,
where
the bosonization technique is available \cite{Giamarch87,Doty92}.
For numerical study of the $d=1$ criticality,
see the article \cite{Hatano95} and references therein.)
Because of the absence of the upper critical dimension,
the $\epsilon$-expansion scheme cannot be formulated.
Hence, for $d=2$, numerical-simulation studies have been playing
a crucial role so far.

Runge \cite{Runge92} employed the exact-diagonalization method to treat
system sizes up to $L=4$; 
the Hamiltonian is given by eq. (\ref{Hamiltonian}) shown afterwards.
(When treating $L=5$, he reduced the particle density
to $n=0.2$.)
He obtained the estimates $\nu=1.4\pm0.3$, $z=1.95\pm0.25$ 
and $\sigma_{\rm c}=0.17\pm0.01((2e)^2/h)$, and
claimed that finite-size correction prevents 
definite conclusion.
Monte-Carlo method has been used in many studies.
Makivi{\'c}, Trivedi and Ullah
\cite{Makivic93}  obtained the estimates $\nu=2.2\pm0.2$, $z=0.5\pm0.05$
and $\sigma_{\rm c}=1.2\pm0.2$.
Batrouni, Larson, Scalettar, Tobochnik and Wang
\cite{Batrouni93}
obtained the estimate $\sigma_{\rm c}=0.45\pm0.07$.
Wallin, S{\o}rensen, Girvin and Young 
\cite{Wallin94} obtained $\nu=0.9\pm0.1$,
$z=2.0\pm0.1$ and $\sigma_{\rm c}=0.14\pm0.03$.
S{\o}rensen, Wallin, Girvin and Young
\cite{Sorensen92} obtained $\nu=1.0\pm0.1$,
$z=2.0\pm0.1$ and $\sigma=0.14\pm0.03$.
Zhang, Kawashima, Carlson and Gubernatis 
\cite{Zhang95} obtained 
$\nu=0.9\pm0.1$ and $z=2.0\pm0.4$.
The estimates are still rather scattering.
In particular, computing the conductivity more reliably
would be of practical importance,
because it can be compared with the experimentally observed
value $\sigma_{\rm c}\approx(2e)^2/h$ mentioned above.
The Monte-Carlo method enables one to treat larger systems
than those with the exact-diagonalization method.
In the quantum Monte Carlo simulation, however,
because the dynamical critical exponent is predicted to be $z=2$,
in order to simulate ground-state property,
the imaginary-time system size should be enlarged {\em rapidly};
namely, it should be kept quadratic
in terms of the real-space system size at least.

In the present paper, we simulate the two-dimensional bose system
on the $L \times L$ square lattice
under the periodic boundary condition
in the presence of the quenched random chemical potential.
The Hamiltonian is given by,
\begin{equation}
{\cal H}=
-\frac{J}{2}\sum_{\langle ij\rangle}
\left( a^\dagger_i a_j + {\rm h.c.} \right)
+
\sum_i H_i(2 a^\dagger_i a_i -1),
\label{Hamiltonian}
\end{equation}
where the operators $\{ a_i,a^\dagger_i \}$ obey the hard-core
boson statistics,
\begin{eqnarray}
& & [a_i,a_j]=[a^\dagger_i,a^\dagger_j]
=[a_i,a^\dagger_j]=0\ \ \ (i \ne j), \nonumber \\
& &
\{a_i,a^\dagger_i\}=1\ \ \ {\rm and}\ \ \ a_i a_i=a^\dagger_i a^\dagger_i=0,
\label{hard_core_boson}
\end{eqnarray}
and $\sum_{\langle ij \rangle}$ denotes the summation over 
all nearest neighbors.
The site-random chemical potentials
$\{ H_i \}$ distribute uniformly over the range
$[-\sqrt{3}\Delta,\sqrt{3}\Delta]$;
namely, the mean deviation is given by 
$\sqrt{[ H_i^2]_{\rm av}}=\Delta$.
The particle density $n=N/L^2$ is fixed to be one half throughout this paper.

As is mentioned above, the model (\ref{Hamiltonian}) is
believed to describe the physics of the superconductivity-localization
transition as well as the superfluid-localization transition.
This belief is 
based on the picture \cite{Ma86,Fisher89}
that the boson describes the Cooper pair, so that the boson
charge should be the twice of the single electron charge; $e^*=2e$.
The model (\ref{Hamiltonian})
is investigated very extensively in the above mentioned article
\cite{Runge92}.
Here, we attempt to improve this work through
utilizing some recent developments,
the algorithm for computing the response function (resolvent)
\cite{Gagliano87},
the estimate scheme of the dynamical critical exponent
\cite{Rieger96},
and parallel supercomputers which enable one to treat larger
system ($L=5$).
These improvements are described in respective subsections
of Section \ref{section3}.

The rest of the paper is organized as follows:
In the next section, we review some
notions relevant to
the present study: 
First, we show the viewpoint from which
the model (\ref{Hamiltonian}) is regarded as the quantum $XY$ model
\cite{Matsubara56}.
Nature of the superfluid-localization transition is interpreted
in the language of the quantum spin system.
Then, we summarize the scaling argument \cite{Fisher89,Fisher90}
 describing
the criticality.
The argument yields some scaling formulae
which are useful in analyzing
our finite-size numerical data.
In the section \ref{section3}, our numerical simulation results are presented.
We estimated the critical exponents as $\nu=2.3\pm0.6$ and $z=2$.
At the critical point, we estimate the conductivity
as $\sigma_{\rm c}=0.135\pm0.01((2e)^2/h)$.
These new results are summarized in the last section
in comparison with previous results.

\section{Review --- equivalence to the $XY$ model and scaling
argument}
\label{section2}

In this section, we summarize several important aspects
about the model (\ref{Hamiltonian}).
First, we introduce the equivalence between the model (\ref{Hamiltonian})
and the quantum $XY$ spin system \cite{Matsubara56}. 
This equivalence reveals nature of both the superfluid and
the localization phases in the language of the spin system,
and provides intuitive picture of the phase transition.
Finally, we review the scaling argument \cite{Fisher89,Fisher90},
whose formulae are
used in the analyses of our numerical data in Section \ref{section3}.

\subsection{Mapping to the quantum $XY$ model with the random 
magnetic field}
\label{section2_1}

Because the $S=1/2$ ladder operators $\{S^+_i,S^-_i\}$ obey the same algebra
 eqs. (\ref{hard_core_boson}) \cite{Matsubara56}, 
the Hamiltonian (\ref{Hamiltonian}) is expressed in terms
of the spin operators,
\begin{equation}
{\cal H}=-J\sum_{\langle ij \rangle}
(S^x_i S^x_j + S^y_i S^y_j)
+2\sum_i H_i S^z_i.
\label{Heisenberg_Hamiltonian}
\end{equation}

Now, perspectives developed for the quantum spin system 
become available.
Kishi and Kubo \cite{Kishi89}
showed rigorously that (without the random 
magnetic field)
in the thermodynamic limit,
the ground-state magnetism is spontaneously broken;
\begin{equation}
\langle {\bf m}_{XY}^2 \rangle \ne 0,
\label{Kubo_Kishi}
\end{equation}
where ${\bf m}_{XY}=\frac{1}{L^2}\sum_i(S^x_i,S^y_i)$.
It is expected that for sufficiently strong random fields,
the long-range magnetism would be disturbed, 
$\langle {\bf m}_{XY}^2\rangle=0$.
Hence, we see that a phase transition
between the $XY$ and the random-field phases would exist
at a certain random-field strength.

What does the existence of the $XY$ order (\ref{Kubo_Kishi}) stand for 
in the boson language?
We show that it stands for
the bose condensation (superfluidity):
If the number of the particles condensing at the ${\bf k}=0$ state
is of the system-size order, the state is regarded as being
of superfluid.
That is, the quantity,
\begin{equation}
|\Psi_{\rm super}|^2=
\frac{1}{L^2}
\left.
\left\langle 
\left( 
\frac{1}{L}\sum_{\bf j} {\rm e}^{{\rm i}{\bf k}\cdot{\bf j}} 
                                  a^\dagger_{\bf j} \right)
\left( 
\frac{1}{L}\sum_{\bf j} {\rm e}^{-{\rm i}{\bf k}\cdot{\bf j}}a_{\bf j} \right)
\right\rangle
\right|_{{\bf k}=\vec{0}},
\end{equation}
 works as an order parameter of the superfluidity.
The order parameter is expressed in term of the spin language,
$|\Psi_{\rm super}|^2=\langle {\bf m}_{XY}^2 \rangle$.
As is explained above, this remaines finite.
And so, the superfluidity develops actually at the ground state of
the Hamiltonian (\ref{Hamiltonian}) at $\Delta=0$.
It is quite natural that
the gauge degree of freedom,
which is spontaneously broken in the superfluid phase,
is related  to the in-plain spin rotator.
As is introduced in Section \ref{section1},
the essence of the superfluid-localization transition
is believed to be concerned only in the in-plain rotator degrees of freedom,
and the site-random perturbations conjugate to them.

\subsection{Scaling argument}
\label{section2_2}

Here, we introduce a scaling argument 
\cite{Fisher89,Fisher90} which describes
the criticality of the superfluid-localization transition.
The argument yields various useful formulae to
analyze numerical data.
The argument itself, however, does not yield any conclusions
to estimate the critical exponents quantitatively.
Any analytical analyses to estimate the critical exponents
face a difficulty as is mentioned in Introduction.
For the purpose of estimating the exponents quantitatively,
numerical simulation has been playing a crucial role.

The scaling hypothesis states that (the singular part of)
the free energy per unit volume for the system size $L$ and
the inverse temperature $\beta$ should be given in the form,
\begin{equation}
f(L,\beta)\sim
\frac{1}{\xi_{\rm r}^d\xi_\tau}
\tilde{f}
\left(
\frac{L}{\xi_{\rm r}},
\frac{\beta}{\xi_\tau}
\right),
\label{FFS_kasetu}
\end{equation}
where $\xi_{\rm r}$ and $\xi_\tau$ denote the real-space correlation length 
and the imaginary-time one, respectively.
(Note that in the path-integral viewpoint,
the partition function of a $d$-dimensional quantum system
is regarded as that of a $(d+1)$-dimensional counterpart.)
These are expressed in term of the deviation from the critical point
$\delta$; $\xi_{\rm r}\sim\delta^{-\nu}$ and 
$\xi_\tau\sim\xi_{\rm r}^z$.
The parameter $\nu$ denotes the correlation-length critical exponent,
and $z$ denotes the dynamical critical exponent.
These are estimated in Section \ref{section3}, numerically.
In general, for quantum random systems,
the real-space correlation develops less robustly than
the imaginary-time one does.
This anisotropy is one of the significant characteristics of the
random quantum critical phenomena, resulting in
various new exotic universality classes.
The anisotropy is characterized by the dynamical critical
exponent $z$.

The superfluid density $\rho_{\rm s}$ ---
the spin stiffness in the language of the spin system ---
is defined as the elastic constant
in terms of the real-space gauge twist \cite{Fisher73},
\begin{equation}
f\sim\frac{\rho_{\rm s}}{2}(\partial_x \theta)^2.
\end{equation}
Note that the quantity works 
as an order parameter of the superfluidity. 
It is furthermore expressed in the form,
$f\sim\frac{\rho_{\rm s}}{2}\left(\frac{\Theta}{L}\right)^2$, 
where $\Theta$ denotes the total gauge twist.
Using the form (\ref{FFS_kasetu}), we obtain,
\begin{eqnarray}
\rho_{\rm s} &\sim& f\cdot L^2                   \nonumber \\
\label{SF_scaling}
             &=& \xi_{\rm r}^{-(d+z-2)}\tilde{\widetilde{f}}
                   \left(\frac{L}{\xi_{\rm r}},\frac{\beta}{\xi_\tau}\right)
                                                           \\
             &=& L^{-(d+z-2)}\tilde{\tilde{\tilde{f}}}
                   \left(\frac{L}{\xi_{\rm r}},\frac{\beta}{\xi_\tau}\right)
                                                 \nonumber
\end{eqnarray}
This scaling form is used in our numerical-data analysis.
The compressibility $\kappa$ is defined, on the other hand,
as the elastic constant of the imaginary-time gauge twist,
$f\sim\frac{\kappa}{2}(\partial_\tau\theta)^2$.
Through the similar arguments as the above,
we obtain the scaling formula for the compressibility,
\begin{equation}
\kappa\sim\xi_{\rm r}^{-(d-z)}\tilde{\tilde{f}}
\left(\frac{L}{\xi_{\rm r}},\frac{\beta}{\xi_\tau}\right).
\end{equation}
Fisher, Weichman, Grinstein and Fisher
predicted that the formula,
\begin{equation}
z=d,
\label{FWGF_yosou}
\end{equation}
should hold
for any dimensions:
The compressibility is finite in both phases beside the transition point.
Hence, it would be kept to be of the order unity
even at the critical point as well,
so that we obtain the equality (\ref{FWGF_yosou}).
In Section \ref{section3}, we confirm their prediction with use of the
Rieger-Young method \cite{Rieger96}.

Finally, we explain how the above scaling argument
concludes that the conductivity would be finite at the
critical point.
The scaling argument \cite{Fisher90} yields the following formula
for the AC conductivity,
\begin{equation}
\sigma(\omega)\sim
\xi_{\rm r}^{-(d+z-2)}\tilde{\tilde{\rho_{\rm s}}}(\omega\xi_\tau)/\omega.
\end{equation}
Assuming $d=2$, and the scaling function
behaves as $\tilde{\tilde{\rho_{\rm s}}}(x)\propto x$ 
in the vicinity of the critical point $x\to\infty$,
we see that
the conductivity does not have any singularities
at the critical point, and therefore
remains 
{\em finite}.
To summarize, the scaling argument states that there is a
possibility that the conductivity remains finite at the
superfluid-localization critical point.
The quantitative evaluation of $\sigma_{\rm c}$,
however, lies out of the scope of the argument.
Here, we compute the conductivity with use of the
Gagliano-Balseiro method \cite{Gagliano87} in the next section.

\section{Numerical results}
\label{section3}

In this section, we present our numerical results.
In order to diagonalize the Hamiltonian (\ref{Hamiltonian}),
we employed the Lanczos method.
We have fixed the
particle density to be one half $n(=N/L^2)=0.5$,
and treated the system sizes
up to $L=5$.
For those system sizes of odd $L$, we proceeded the sets of
simulations for the particle numbers $N=[L^2/2]$ and $[L^2/2]-1$;
the bracket $[\cdots]$ denotes the Gauss notation.
The data for $n=0.5$ are obtained through
interpolating these two sets of data with use of the relation
$Q(n)=a(n-0.5)^2+c$;
physics is symmetric in terms of $n=0.5$ (particle-hole symmetry).

\subsection{Criticality of the superfluid-localization transition}
\label{section3_1}

Here, we determine the location of the 
superfluid-localization transition point and the correlation-length
critical exponent.
We use the language of the spin system, which is explained
in the  subsection \ref{section2_1}.
That is, from this viewpoint,
the transition is characterized by the disappearance of
the $XY$ (in-plain) magnetic order.

\begin{figure}[htbp]
\begin{center}\leavevmode
\epsfxsize=8.5cm
\epsfbox{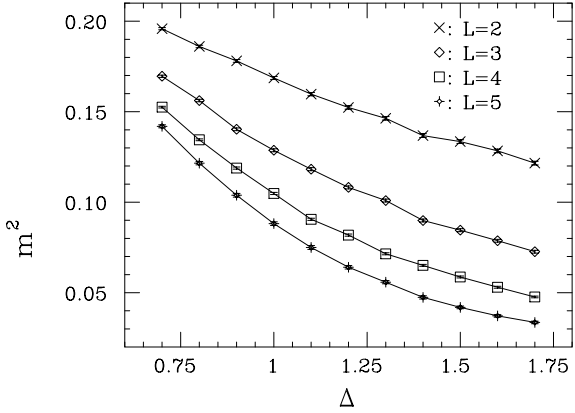}
\end{center}
\caption{Square of the in-plain magnetization $m^2$
is plotted
against the randomness $\Delta$.
The random-sample numbers are $2048$, $2048$, $2048$ and $608$ for
$L=2$, $3$, $4$ and $5$, respectively. 
The $XY$ magnetic order becomes suppressed as the
random magnetic field is strengthened.}
\label{magnetization}
\end{figure}

\begin{figure}[htbp]
\begin{center}\leavevmode
\epsfxsize=8.5cm
\epsfbox{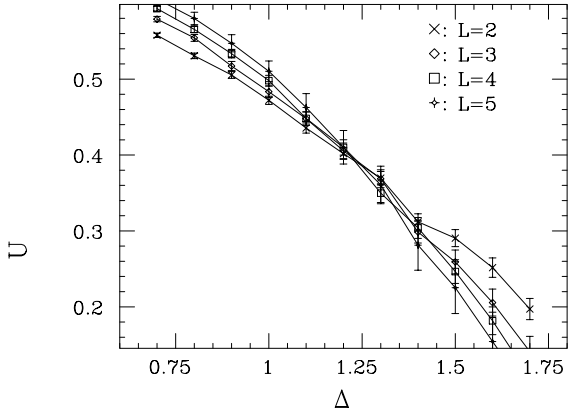}
\end{center}
\caption{Binder parameter 
({\protect \ref{Binder_parameter}}) of the in-plain magnetization
is plotted against the randomness.
The intersection point of the curves indicates the location
of the transition point.}
\label{Binder}
\end{figure}

In Fig. \ref{magnetization}, we plotted the square of the in-plane
magnetization 
$m^2=[\langle {\bf m}_{XY}^2 \rangle]_{\rm av}$ against the randomness,
where the bracket $[\cdots]_{\rm av}$ denotes the random-sampling
average, and $\langle \cdots \rangle$ denotes the ground-state
expectation value.
The random-sample numbers are $2048$, $2048$ $2048$ and $608$ for
$L=2$, $3$, $4$ and $5$, respectively. 
We see that the in-plane magnetization becomes suppressed
by the randomness.
In Fig. \ref{Binder}, we plotted the Binder parameter 
\cite{Binder81} of the
in-plane magnetic order,
\begin{equation}
U=1-\frac{[\langle {\bf m}_{XY}^4 \rangle]_{\rm av}}
{3[\langle {\bf m}_{XY}^2 \rangle^2 ]_{\rm av}}.
\label{Binder_parameter}
\end{equation}
The Binder parameter is invariant with respect to
the system sizes at the critical point.
It is enhanced (suppressed) in the order (disorder) region
as the systems size is enlarged.
In Fig. \ref{Binder},
We observe an intersection point at $\Delta\approx1.2$.
Namely, in the region $\Delta<1.2$,
the superfluidity persists against the random chemical potential,
whereas in $\Delta>1.2$, the particles are localized,
and the long-range gauge coherence is lost.

In order to estimate the transition point and the 
correlation-length exponent precisely,
we analyze the above data by means of the finite-size-scaling 
theory.
According to the theory, the Binder parameter (dimensionless
quantity) obeys the form 
$U=\tilde{U}\left((\Delta-\Delta_{\rm c})L^{1/\nu}\right)$ 
in the vicinity of the
critical point $\Delta_{\rm c}$.
Namely, the data $(\Delta-\Delta_{\rm c})L^{1/\nu}$-$U$ should collapse
along a universal curve irrespective of the system sizes.
In other words, the critical point $\Delta_{\rm c}$ and the exponent $\nu$
are adjusted so that the scaled data could form a universal curve.
The degree to what extent the data collapse is measured by
the Kawashima-Ito ``local-linearity function''
\cite{Kawashima93} which is explained 
in Appendix.
In Fig. \ref{scaling}, we show the scaling plot for the data
shown in Fig. \ref{Binder}.
\begin{figure}[htbp]
\begin{center}\leavevmode
\epsfxsize=8.5cm
\epsfbox{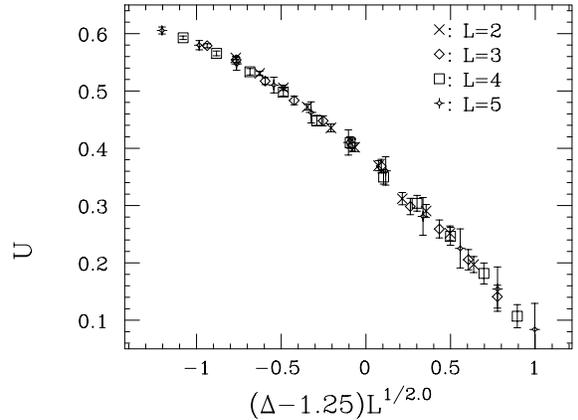}
\end{center}
\caption{Scaling plot for the data shown in
Fig. {\protect \ref{Binder}}.
The scaling analysis yields the best-fit
estimates $\Delta_{\rm c}=1.25$
and $\nu=2.0$.}
\label{scaling}
\end{figure}
In consequence,
we obtain the estimates
$\Delta_{\rm c}=1.25\pm0.1$ and $\nu=2.3\pm0.6$.
In order to see whether the correction to the finite-size scaling
exists, we analyzed the data for $L=3$, $4$ and $5$ similarly;
omitted the data for $L=2$.
This analysis yields the best-fit
estimates $\Delta_{\rm c}=1.27$ and $\nu=2.3$.
Therefore, we conclude that there is little correction to the finite-size
scaling;
in other words, the system sizes treated here reach the scaling
region.
Our estimate $\nu=2.3\pm0.6$ is somewhat different
from that of ref. \cite{Runge92}
($\nu=1.4\pm0.3$)
despite of the fact that we considered the same model
also using the exact-diagonalization technique.
This discrepancy may originate in the
criteria utilized to appreciate the scaling-data collapse,
and is discussed in the next section.

\begin{figure}[htbp]
\begin{center}\leavevmode
\epsfxsize=8.5cm
\epsfbox{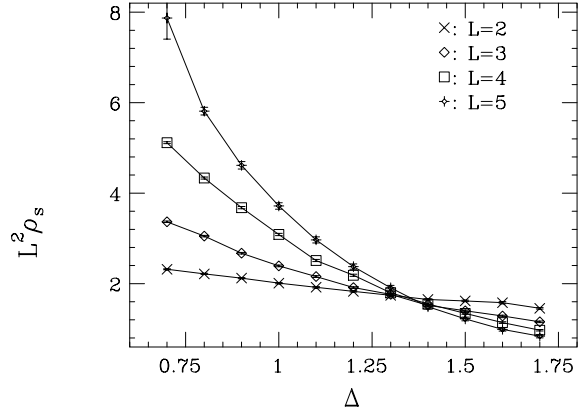}
\end{center}
\caption{Scaled spin stiffness $L^2\rho_{\rm s}$ is plotted.
Because the curves intersect at a point, the prediction
$z=d$ is appeared to hold.}
\label{stiffness}
\end{figure}

According to the formula (\ref{SF_scaling}), if we assume that
the dynamical critical exponent is equal to the spacial dimension,
namely, $z=d=2$, the scaled spin stiffness
$L^2\rho_{\rm s}$ should be invariant at the critical point.
The spin stiffness is given by,
\begin{equation}
\rho_{\rm s}=
\left[
\left\langle
\left.
\frac{\partial^2 E_{\rm g}(\Theta)}{\partial \Theta^2}
\right|_{\Theta=0}
\right\rangle
\right]_{\rm av} \ \ ,
\end{equation}
where the angle denotes the boundary gauge twist
$a^\dagger_i a_j \to {\rm e}^{{\rm i}\Theta} a^\dagger_i a_j$,
and $E_{\rm g}$ denotes the ground-state energy.
In Fig. \ref{stiffness}, we plotted the scaled spin stiffness
$L^2\rho_{\rm s}$.
In fact, we observe at $\Delta\approx1.3$, all the curves intersect.
In consequence, founded on the scaling formula (\ref{SF_scaling}),
the so-called ``generalized Josephson relation,''
we see that the dynamical exponent 
would be equal to two.
The same reasoning founded on this relation
has been reported in the literatures 
\cite{Runge92,Makivic93,Wallin94,Zhang95}.
In this paper, 
in Section \ref{section3_3},
we estimate the dynamical critical exponent with use
of the Rieger-Young \cite{Rieger96}
method, which would be more straightforward.

\subsection{Electrical conductivity}
\label{section3_2}

We evaluate the dynamical conductivity at the critical point 
$\Delta_{\rm c}=1.25$
estimated in the above subsection.
The DC conductivity is conjectured
to be universal at the critical point as 
is explained in Introduction.
The dynamical conductivity is given by the 
current-current time correlation,
\begin{equation}
\sigma(\omega)={\rm Re}
\left[
\frac{1}{\hbar\omega}\frac{1}{L^2}\int_0^\infty {\rm d}t
{\rm e}^{{\rm i}\omega t}
\langle[J_x(t),J_x]\rangle
\right]_{\rm av},
\label{Kubo_formula}
\end{equation}
where the current is given by 
$J_x=\frac{{\rm i}e^*J}{2\hbar}\sum_{j,\delta_x}\delta_x 
a^\dagger_{j+\delta_x} a_j$.
In the previous work \cite{Runge92}, the conductivity is calculated 
through Fourier-transforming the current-current correlation numerically,
whose time correlation is
evaluated with use of the Suzuki-Trotter-%
decomposition approximation.
We evaluate the resolvent form of eq. (\ref{Kubo_formula}) directly,
\begin{eqnarray}
& & \sigma(\omega)={\rm Re}\left(\frac{\rm i}{\hbar\omega L^2}
                   \right.  \nonumber \\
& & \times \left.
\left[
\left\langle
J_x
\left(
\frac{\hbar}{E_{\rm g}-{\cal H}+\hbar\omega}+
\frac{\hbar}{E_{\rm g}-{\cal H}-\hbar\omega}
\right)
J_x
\right\rangle
\right]_{\rm av}
\right).
\label{resolvent}
\end{eqnarray}
Now, we are free from the discrete numerical integration error.
(The conductivity is a linear combination of the delta-function peaks
as is apparent from eq. (\ref{resolvent}),
which might be suffered significantly from the 
numerical discretization error.)
Some might wonder that the inverse matrix of the total Hamiltonian
in eq. (\ref{resolvent})
cannot be computed; this is true.
The {\em expectation value} of the inverse of the Hamiltonian is,
however, evaluated with use of the Gagliano-Balseiro
continued-fraction formula \cite{Gagliano87},
\begin{equation}
\left\langle f_0 
\left| \frac{1}{z-{\cal H}} 
\right| f_0 \right\rangle
=
\frac{\langle f_0 | f_0 \rangle}
{
z-\alpha_0-\frac{\beta_1^2}
{
z-\alpha_1-\frac{\beta_2^2}{\ddots}
}
},
\label{Gagliano-Balseiro}
\end{equation}
where the coefficients are given by the Lanczos tri-diagonal
elements,
\begin{eqnarray}
|f_{i+1}\rangle &=& {\cal H}|f_i\rangle-\alpha_i|f_i\rangle
                      -\beta_i^2|f_{i-1}\rangle,           \\
\alpha_i &=& \langle f_i|{\cal H}|f_i\rangle/\langle f_i|f_i\rangle , 
                                                 \nonumber \\
\beta_i^2 &=& \langle f_i|f_i\rangle/\langle f_{i-1}|f_{i-1}\rangle
\ \ \   (\beta_0 = 0). \nonumber
\end{eqnarray}
We have evaluated the dynamical conductivity (\ref{resolvent}) 
by means of this formula.

\begin{figure}[htbp]
\begin{center}\leavevmode
\epsfxsize=8.5cm
\epsfbox{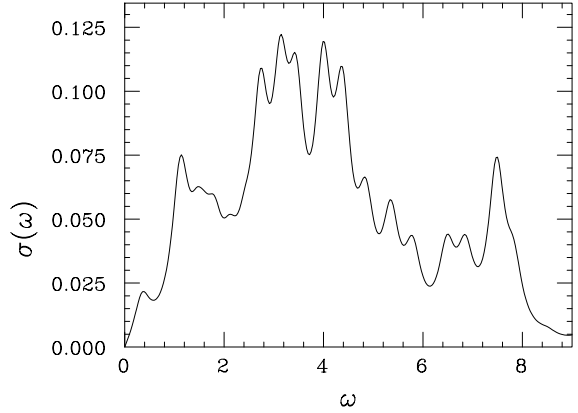}
\end{center}
\caption{Electrical conductivity is evaluated for a random sample
with $\Delta=1.25$ and $L=5$.
We see that the conductivity consists of delta-function peaks;
the delta-function is broadened into the Lorentz form with
the width $\eta=0.2$.}
\label{dendou_sample}
\end{figure}

In Fig. \ref{dendou_sample}, we show the conductivity for a certain 
random sample
with $\Delta=1.25$ and $L=5$.
We see that the conductivity consists of delta-function peaks.
The delta function peak is broadened
into the Lorentz form through the
substitution $\omega\to\omega-0.2{\rm i}$.
The concept of the dissipation becomes
subtle for finite-size system.
For instance, the DC
conductivity is vanishing.
This is due to the presence of the finite-size
energy gap above the ground state.
Only in the thermodynamic limit,
The conductivity in the vicinity of the zero frequency $\omega=0$
would emerge. 
(In the quantum Monte Carlo simulation,
the (real-)time correlation cannot be computed;
the temperature (imaginary time) correlation function
is computed instead.
The latter is free from the delta-function singularities,
because the poles exist only along the real-frequency axis.
An analytical continuation to the real frequency, however,
should be performed in order to estimate the
dynamical response function such as eq. (\ref{Kubo_formula}).
This is extremely difficult.)

\begin{figure}[htbp]
\begin{center}\leavevmode
\epsfxsize=8.5cm
\epsfbox{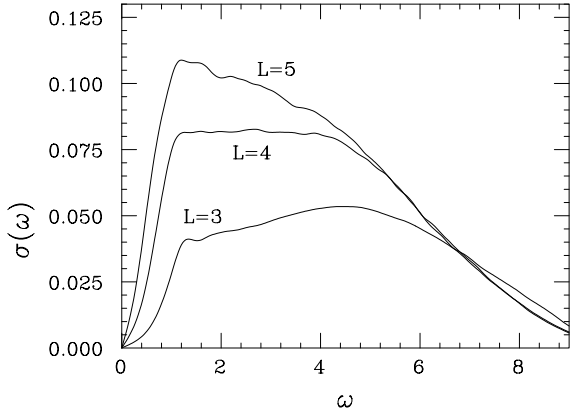}
\end{center}
\caption{Electrical AC conductivity is evaluated with
the Gagliano-Balseiro formula ({\protect \ref{Gagliano-Balseiro}}) for
$\Delta=1.25$; the random-sample numbers are
$16384$, $2048$ and $960$ for $L=3$, $4$ and $5$, respectively.}
\label{dendou}
\end{figure}

\begin{figure}[htbp]
\begin{center}\leavevmode
\epsfxsize=8.5cm
\epsfbox{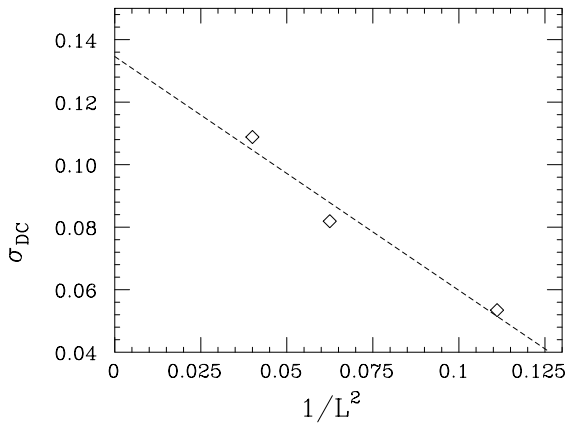}
\end{center}
\caption{$1/L^2$ extrapolation of the DC conductivity.}
\label{gaisou}
\end{figure}

In Fig. \ref{dendou}, we show the random-averaged conductivity for
$\Delta=1.25$;
the random-sample numbers are
$16384$, $2048$ and $960$ for $L=3$, $4$ and $5$,
respectively.
We see that the conductivity increases as the frequency is reduced.
In the vicinity of the static point $\omega\sim0$, however,
the conductivity drops rapidly due to the reason mentioned above
(finite-size effect).
(For comparison, we show the Fourier-transformed
results \cite{Runge92};
$\sigma\approx0.055$ for $L=3$ and $\sigma\approx0.105$ for $L=4$.
These are a bit larger than ours.)
We estimate the DC conductivity as the maximal value
of the AC conductivity.
The DC conductivity shows large system-size dependence.
In Fig. \ref{gaisou}, we depict the $1/L^2$ extrapolation
of the conductivity.
(The power $2$ is chosen for the same reasoning as that in the paper
\cite{Runge92}.)
The plots align.
We stress that the present new data for $L=5$ is crucial to
confirm the validity of the $1/L^2$ extrapolation.
We obtained the extrapolated conductivity with
the least-square fit as $\sigma_{\rm c}=0.135\pm0.01
((2e)^2/h)$.

\subsection{Dynamical critical exponent}
\label{section3_3}

According to the conjecture introduced in Section \ref{section2_2},
the dynamical critical exponent 
is given by $z=2$ at the superfluid-localization transition.
This conjecture has been confirmed numerically 
\cite{Runge92,Wallin94,Zhang95} with the help
of the generalized Josephson relation (\ref{SF_scaling}) 
for the superfluid density.
In fact, we demonstrated in Section \ref{section3_1} that 
the superfluid density is well described in terms of the
generalized Josephson relation and the assumption $z=2$.
In this subsection, we utilize the Rieger-Young formula
\cite{Rieger96} for the first time
in order to estimate $z$.
We think that the scheme is suitable for the
exact-diagonalization simulation, and gives $z$ straightforwardly.

\begin{figure}[htbp]
\begin{center}\leavevmode
\epsfxsize=8.5cm
\epsfbox{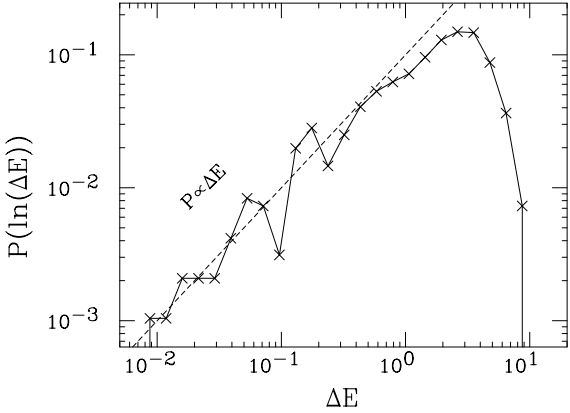}
\end{center}
\caption{Probability distribution of the first energy gap
for the system with $\Delta=1.25$ and $L=5$.}
\label{bunpu}
\end{figure}

In Fig. \ref{bunpu}, we show the probability distribution of the
first energy gap $\Delta E$;
$\Delta=1.25$, $L=5$ and 960 random samples.
According to the Rieger-Young argument, from the low-energy tail
of the distribution,
the dynamical critical exponent is extracted:
The probability of a certain energy gap
may be proportional to the spatial volume
$P \propto L^d\cdot\Delta E^\lambda$,
where the exponent $\lambda$ describes the low-energy tail.
On the other hand, the finite-size scaling theory states that
any quantities should be expressed in the form 
$P=\tilde{P}(L/\xi_{\rm r})=\tilde{P}(L/\xi_\tau^{1/z})$;
see the subsection \ref{section2_2}.
With use of the formula $\Delta E\sim1/\xi_\tau$,
the distribution turns out to be a function of $L(\Delta E)^{1/z}$.
Through collating this fact with the above form,
we obtain the relation $\lambda=d/z$, which is the Rieger-Young relation.

In Fig. \ref{bunpu}, we see that the low-energy tail
is, in fact, well described by the exponent $\lambda=1$;
namely, we obtain the estimate $z=d=2$.

\section{Summary and discussions}
\label{section4}

We have investigated the two-dimensional hard-core boson system
in the presence of the random chemical potential,
whose Hamiltonian is given by (\ref{Hamiltonian}).
We diagonalized the Hamiltonian numerically for the system sizes 
up to $L=5$.
The present new data for $L=5$ show no unexpected (irregular)
behavior in the finite-size-scaling analysis.
That is, the system sizes treated here reach scaling
regime.
This was not so certain, and in fact worried about in the
previous study.
We observed that the disorder-driven superfluid-localization transition
takes place at the critical randomness $\Delta_{\rm c}=1.25\pm0.1$.
We estimated the correlation-length exponent
and the dynamical critical exponent as $\nu=2.3\pm0.6$
and $z=2$, respectively,
with the finite-size scaling analysis.
By means of the Gagliano-Balseiro method,
we computed the dynamical conductivity at the critical point.
Thereby, we obtained the DC conductivity 
$\sigma_{\rm c}=0.135\pm0.01 ((2e)^2/h)$.

Our estimate of the critical point is accordant with that of Runge
\cite{Runge92}:
By means of the
same numerical method as ours,
Runge obtained $\Delta_{\rm c}\sim1.15$ through the
analysis of $\rho_{\rm s}$, $\Delta_{\rm c}\sim1.25$ through the
Binder parameter (the same as ours) and $\Delta_{\rm c}\sim1.1$-$1.3$
through $\langle{\bf m}_{XY}^2\rangle$.
The present new simulation for larger system size ($L=5$)
confirmed that 
the system size is surely in the scaling regime.
On the other hand, 
by means of the quantum Monte Carlo method
it has been reported that the transition point locates
at $\Delta_{\rm c}=1.43\pm0.06$ \cite{Zhang95} and $0.72$ \cite{Makivic93}
(in our definition).
These estimates are based on the generalized Josephson relation
(\ref{SF_scaling}) and an assumption of $z$;
the former assumed $z=2$, whereas the latter assumed $z=0.5$.
We think that the Binder parameter of the gauge order 
(\ref{Binder_parameter}),
which is readily 
computed with the exact-diagonalization scheme,
shows less correction to the finite-size scaling.



Secondly, we discuss the correlation-length exponent $\nu$.
It should be made clear
why our estimate $\nu=2.3\pm0.6$ differs
from that of ref. \cite{Runge92} ($\nu=1.4\pm0.3$) despite of
the fact that we considered the same model by means of the 
same numerical technique except for our extension to $L=5$.
Crucial difference might be the criteria utilized
to appreciate
the data collapse in the scaling analysis.
As is explained in Appendix, we used a quantitative scheme
which also gives the estimate of the error margin.
The error margin is, however, determined through
taking account of only the statistical error, but not
the correction to finite-size scaling.
The latter
is hardly appreciable quantitatively.
Hence, there is a possibility that the error margin would be
somewhat larger.
The estimate of ref. \cite{Runge92}
lies a bit outside of our error bar,
and, in fact,
the scaling plot based on the assumption $\nu=1.4$
does not show distinct failure as far as one can see.
We believe, however, that our quantitative criterion would be 
objective and
possibly less biased, although the present error margin
might be rather large.
At least,
our data shown in Fig. \ref{linearity} seem to exclude
the conclusion $\nu\sim1$.
In fact, Chayes {\it et al.} claimed an inequality $\nu\ge2/d$,
\cite{Chayes86,Chayes89}
which casts doubt on $\nu<1$.


Next, we turn to discuss the dynamical critical exponent.
We have used the Rieger-Young relation for the first time
to estimate the dynamical critical exponent.
The analysis showed a clear evidence
that $z(=d)=2$ holds.
Because the previous estimates founded on the generalized
Josephson relation (\ref{SF_scaling}) indicate the same conclusion as well,
we believe that
the prediction $z=d$ is established fairly definitely
in two dimensions.

Finally, we mention about the electrical conductivity.
The present conclusion $\sigma_{\rm c}/((2e)^2/h)=0.135\pm0.01$ is 
comparable with the estimates
$0.17\pm0.01$ \cite{Runge92} and $0.14\pm0.03$ \cite{Wallin94}.
We stress that the previous exact-diagonalization data \cite{Runge92}
for $L=3$ and $4$ are rather far
from convergence, and our result for $L=5$ reveled that
the result of the system size is about to converge 
to a certain thermodynamic-limit value.
In fact, as is shown in Fig. \ref{gaisou},
our new data $L=5$ is vital in order to confirm the validity
of the $1/L^2$ extrapolation.
Quantum Monte Carlo method is less efficient 
in computing dynamical quantity.
(In the quantum Monte Carlo simulation,
the temperature (imaginary time) correlation
rather than the time correlation is evaluated instead.
The temperature correlation is fitted by
an analytical function so as to yield the time correlation
through the Wick rotation.)
The present calculation with the help of the
Gagliano-Balseiro method compensates the disadvantage.
 
In the paper \cite{Damle97}, the authors claim that 
the limit $\omega\to0$ at $T=0$,
which is used in the present paper, fails to give
the DC conductivity  relevant to the experimental transport measurements.
According to their theory,
the AC conductivity shows a Drude-like peak with 
the width $k_{\rm B}T/\hbar$.
Hence, at $T\to0$, the peak vanishes to a single point
so as to make the AC conductivity singular (discontinuous) at $\omega=0$.
Through taking account of this, the authors succeeded in
explaining the experimentally observed critical 
conductivity $\approx(2e)^2/h$.
Our numerical simulation is rather incapable of 
observing the features they proposed.
This remained to be solved in future.

\section*{Acknowledgement}
Our program is based on the subroutine package TITPACK ver. 2
coded by professor H. Nishimori.
Numerical simulations were performed on VPP/700 56
of the computer center, Kyushu university.

\appendix

\section{Details of the present scaling analyses}
\label{appendix}

We explain the details of our finite-size-scaling analyses,
which we managed in Section \ref{section3}
in order to estimate the transition point $\Delta_{\rm c}$ and
the exponent $\nu$.
We adjusted these scaling parameters
so that the scaled data shown in
Fig. \ref{scaling}
form a universal curve irrespective of the system sizes.
In order to see quantitatively to what extent these data align,
we employ the ``local linearity function'' $S$
defined by Kawashima and Ito \cite{Kawashima93}:
Suppose
a set of the data points $\{ (x_i,y_i) \}$ with the error-bar
$\{ d_i(=\delta y_i) \}$, which
we number so that $x_i<x_{i+1}$ may
hold for $i=1,2,\cdots,n-1$.
For this data set, the local-linearity function is defined as
\begin{equation}
S=\sum_{i=2}^{n-1} 
w(x_i,y_i,d_i|x_{i-1},y_{i-1},d_{i-1},x_{i+1},y_{i+1},d_{i+1}).
\label{linear}
\end{equation}
The quantity $w(x_j,y_j,d_j|x_i,y_i,d_i,x_k,y_k,d_k)$
is given by
\begin{equation}
w=
\left(
\frac
{y_j-\bar{y}}
{\Delta}
\right)^2 \ ,
\end{equation}
where
\begin{equation}
\bar{y}=\frac{(x_k-x_j)y_i-(x_i-x_j)y_k}
{x_k-x_i}
\end{equation}
and
\begin{equation}
\Delta^2=d^2_j+
\left(
\frac{x_k-x_j}{x_k-x_i} d_i
\right)^2
+
\left(
\frac{x_i-x_j}{x_k-x_i}d_k
\right)^2.
\end{equation}
In other words, the numerator $y_j-\bar{y}$ 
denotes the deviation of the point 
$(x_j,y_j)$
from the line passing
two points $(x_i,y_i)$ and $(x_k,y_k)$, and
the denominator $\Delta$ stands for the statistical error
of $(y_i-\bar{y})$.
And so, $w=((y_i-\bar{y})/\Delta)^2$ shows a degree to what extent
these three points align.
The advantage in the above analysis is as follows:
In the conventional least square fitting, we need to assume
some particular fitting function. The assumption
which function we use causes systematic error.
Note that in the present analysis, we do not have to assume
any fitting functions.

\begin{figure}[htbp]
\begin{center}\leavevmode
\epsfxsize=8.5cm
\epsfbox{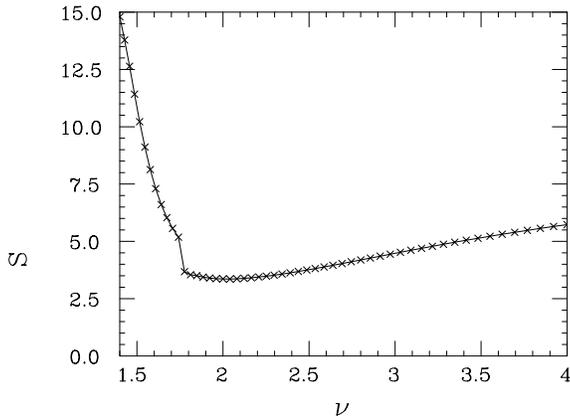}
\end{center}
\caption{Local linearity function is plotted with varying
the scaling parameter $\nu$.
From this plot, we estimated the exponent as $\nu=2.3\pm0.6$.}
\label{linearity}
\end{figure}

The scaling parameters are determined so as to minimize the
local-linearity function.
The error margin of the scaling parameter is hard to estimate.
It is concerned with both the statistical error and the correction to the
finite size scaling.
The former error can be estimated through considering the 
statistical error of the function $S$.
This function has a relative error of the order $1/\sqrt{n-2}$.
(The local-linearity function is of the order $(n-2)$,
and the error is given by $\sqrt{n-2}$.)
The correction to the finite-size scaling,
on the contrary, is quite difficult to determine.
Here, we consider only the statistical error in order
to estimate the error margins of the scaling parameters.
As the number of the data points $n$ is increased,
the statistical error of $S$ is reduced.
The corrections to the finite-size scaling
might increase instead.
In the present analyses, we used twenty data in the vicinity of the
transition point $\Delta_{\rm c}$.
An example of the plot $S$ is shown in Fig. \ref{linearity}.
We observe the minimum at $\nu=2.0$.
Taking into account of the statistical error,
we estimated the critical exponent as $\nu=2.3\pm0.6$.

\end{document}